\documentclass[twoside]{article}        
\usepackage{fleqn,espcrc2}
                                        \usepackage{amssymb,isolatin1}
\def\IP{{\mathbb P}}    \def\IC{{\mathbb C}}    \def\IR{{\mathbb R}}
\def\IZ{{\mathbb Z}}
\def\ifundefined#1{\expandafter\ifx\csname#1\endcsname\relax}
\usepackage{epic,rotating}
\def\makeYZ(#1,#2,#3){
        \Xaux=#1 \multiply \Xaux by \XYfac \Yaux=#2 \multiply \Yaux by \Yfac 
        \advance\Yaux by -\Xaux
        \Xaux=#1 \multiply \Xaux by \XZfac \Zaux=#3 \multiply \Zaux by \Zfac 
        \advance\Zaux by -\Xaux
}
\def\xyzline(#1,#2,#3)(#4,#5,#6){\makeYZ(#1,#2,#3) \YA=\Yaux \ZA=\Zaux 
        \makeYZ(#4,#5,#6)       \drawline(\YA,\ZA)(\Yaux,\Zaux)
}
        
\def\nocolors{\def\TC##1{} \def\CFigOffset{0}}  
\def\subdef#1{\gdef\globalColor##1{##1}}  \def\CFigOffset{4}       
\def\TC#1{\subdef{cmyk #1}}      \subdef{Black}
\def\red{\TC{0 1 1 0}} \def\blue{\TC{1 1 0 0}} \def\black{\TC{0. 0. 0. 1.}}
\ifundefined{ShowColors} \nocolors \fi  

\def\BP{\begin{picture}} \def\EP{\end{picture}}
\def\TwoDimRP{{
        \newcount\XYfac \newcount\Yfac  \newcount\YA    \XYfac=10 \Yfac=18
        \newcount\XZfac \newcount\Zfac  \newcount\ZA    \XZfac=8  \Zfac=18
        \newcount\Xaux  \newcount\Yaux  \newcount\Zaux  \thicklines
                  \linethickness{2pt} \def\TwoDimPtGrid##1##2{{\matrixput
                      (-\CFigOffset,0)(\Yfac,0){##1}(0,\Zfac){##2}{\circle*5}}}
\def\Aa{\BP(60,60)(-10,-5)\blue
        \xyzline(0,0,0)(0,0,3)\xyzline(0,0,0)(0,3,0)\xyzline(0,0,3)(0,3,0)
        \black  \TwoDimPtGrid44 \EP}
\def\Ab{\BP(60,60)(-10,-5)\black
        \xyzline(0,0,0)(0,1,2)\xyzline(0,0,0)(0,2,1)\xyzline(0,1,2)(0,2,1)
        \black  \TwoDimPtGrid44 \EP}
\def\Ba{\BP(100,60)(-10,-5)\blue        
        \xyzline(0,0,0)(0,4,0)\xyzline(0,0,0)(0,2,2)\xyzline(0,2,2)(0,4,0)
        \black  \TwoDimPtGrid53 \EP}
\def\Bb{\BP(60,60)(-10,-5) \black
        \xyzline(0,0,0)(0,2,0)\xyzline(0,0,0)(0,1,2)\xyzline(0,2,0)(0,1,2)
        \black  \TwoDimPtGrid33 \EP}
\def\Ca{\BP(60,60)(-10,-5) \blue                    \xyzline(0,0,0)(0,0,2)
        \xyzline(0,0,0)(0,2,0)\xyzline(0,2,0)(0,2,2)\xyzline(0,0,2)(0,2,2)
        \black  \TwoDimPtGrid33 \EP}
\def\Cb{\BP(60,60)(-10,-5) \black                   \xyzline(0,0,1)(0,1,0)
        \xyzline(0,2,1)(0,1,2)\xyzline(0,0,1)(0,1,2)\xyzline(0,1,0)(0,2,1)
        \black  \TwoDimPtGrid33 \EP}
\def\Da{\BP(60,60)(-10,-5) \black \xyzline(0,0,0)(0,0,2)\xyzline(0,0,0)(0,2,0)
        \xyzline(0,2,0)(0,2,1)\xyzline(0,0,2)(0,1,2)\xyzline(0,1,2)(0,2,1)
        \black  \TwoDimPtGrid33 \EP}
\def\Db{\BP(60,60)(-10,-5) \black \xyzline(0,0,1)(0,0,0)\xyzline(0,0,0)(0,1,0)
        \xyzline(0,2,1)(0,1,2)\xyzline(0,0,1)(0,1,2)\xyzline(0,1,0)(0,2,1)
        \black  \TwoDimPtGrid33 \EP}
\def\Ea{\BP(80,60)(-10,-5)                          \xyzline(0,0,0)(0,0,2)
        \xyzline(0,0,0)(0,3,0)\xyzline(0,0,2)(0,1,2)\xyzline(0,1,2)(0,3,0)
        \black  \TwoDimPtGrid43 \EP}
\def\Eb{\BP(60,60)(-10,-5)
        \xyzline(0,0,0)(0,1,0)\xyzline(0,2,1)(0,1,2)
        \xyzline(0,0,0)(0,1,2)\xyzline(0,1,0)(0,2,1)
        \black  \TwoDimPtGrid33 \EP}
\def\Fa{\BP(80,60)(-10,-5)                          \xyzline(0,0,0)(0,0,1)
        \xyzline(0,0,0)(0,3,0)\xyzline(0,3,0)(0,1,2)\xyzline(0,0,1)(0,1,2)
        \black  \TwoDimPtGrid43 \EP}
\def\Fb{\BP(60,60)(-10,-5)
        \xyzline(0,1,2)(0,0,0)\xyzline(0,0,0)(0,2,0)\xyzline(0,2,0)(0,2,1)
        \xyzline(0,2,1)(0,1,2)
        \black  \TwoDimPtGrid33 \EP}

\def\SA{\BP(80,60)(-10,-5)
\red    \xyzline(0,0,0)(0,1,2)\xyzline(0,0,0)(0,3,0)\xyzline(0,1,2)(0,3,0)
        \black  \TwoDimPtGrid43 \EP}

\def\SB{\BP(60,60)(-10,-5) 
\red    \xyzline(0,0,0)(0,0,2)\xyzline(0,0,0)(0,2,0)
        \xyzline(0,2,0)(0,1,2)\xyzline(0,0,2)(0,1,2)
        \black  \TwoDimPtGrid33 \EP}

\def\SC{\BP(60,60)(-10,-5) 
\red    \xyzline(0,0,0)(0,0,1)\xyzline(0,0,0)(0,2,0)
        \xyzline(0,2,0)(0,2,1)\xyzline(0,0,1)(0,1,2)\xyzline(0,1,2)(0,2,1)
        \black  \TwoDimPtGrid33 \EP}

\def\SD{\BP(60,60)(-10,-5) 
\red    \xyzline(0,1,0)(0,0,1)\xyzline(0,1,0)(0,2,0)\xyzline(0,0,2)(0,0,1)
        \xyzline(0,0,2)(0,1,2)\xyzline(0,2,0)(0,2,1)\xyzline(0,1,2)(0,2,1)
        \black  \TwoDimPtGrid33 \EP}
\parbox{16cm}
{\begin{center}\unitlength=.8pt      {\Yfac=12 \Zfac=12 \hspace{-3pt}
        \Aa ~ \Ab} ~~~~ \Ba ~ \Bb ~~~~~ \Ca ~ \Cb      \\[3pt]
        \Da ~ \Db ~~~~~ \Ea ~ \Eb ~~~~~ \Fa ~ \Fb       \\[3pt]
        \SA ~~~~~~~ \SB ~~~~~~~ \SC ~~~~~~~ \SD         \end{center}}
}}

\def\CICYspec
{\centerline{\unitlength=2.7pt\begin{picture}(150,20)(10,-5)      
        \def\I##1.##2.{\put(##2,##1){\circle*{.9}}}\put(0,0){\vector(1,0){170}}
        \def\h##1.##2.{\put(##2,##1){\circle{.75}}}\put(0,0){\vector(0,1){10}}
        \def\hlab##1{\put(##1,-1){\line(0,1){2}}} \hlab{10}\hlab{20}\hlab{30}
        \hlab{40}\hlab{50}\hlab{60}\hlab{70}\hlab{80}\hlab{90}\hlab{100}
        \hlab{110}\hlab{120}\hlab{130}\hlab{140}\hlab{150}\hlab{160}
        \put(48.5,-4){\small\bf50}\put(97.5,-4){\small\bf100}
        \put(147.5,-4){\small\bf150}\put(166,-4){$\mathbf{h_{11}}$}
        \put(1,8){$\mathbf{h_{12}}$}    \def\i##1.##2.{\h##1.##2.\I##1.##2.}
\i1.25.\i1.37.\i1.61.\i1.73.\i1.79.\i1.89.\i1.129.\i2.30.\i2.36.\i2.44.
\i2.50.\i2.54.\i2.56.\i2.58.\i2.59.\i2.60.\i2.62.\i2.64.\i2.66.\i2.68.
\i2.70.\i2.72.\i2.76.\i2.77.\i2.78.\i2.80.\i2.82.\i2.100.\i2.112.\i3.23.
\i3.24.\i3.27.\i3.29.\i3.31.\i3.33.\i3.35.\i3.37.\i3.39.\i3.41.\i3.42.
\i3.44.\i3.47.\i3.48.\i3.49.\i3.50.\i3.52.\i3.53.\i3.54.\i3.55.\i3.56.
\i3.58.\i3.60.\i3.61.\i3.62.\i3.64.\i3.68.\i3.70.\i3.80.\i3.101.\i3.113.
\i4.16.\i4.22.\i4.24.\i4.26.\i4.30.\i4.31.\i4.32.\i4.33.\i4.38.\i4.39.
\i4.41.\i4.43.\i4.45.\i4.47.\i4.51.\i4.53.\i4.75.\i4.83.\i5.25.\i5.27.
\i5.102.\i6.20.\i6.24.\i7.22.\i8.14.
\h1.21.\h1.101.\h1.103.\h1.145.\h1.149.\h2.29.\h2.38.\h2.74.\h2.83.\h2.84.
\h2.86.\h2.90.\h2.92.\h2.95.\h2.102.\h2.106.\h2.116.\h2.120.\h2.122.\h2.128.
\h2.132.\h2.144.\h2.272.\h3.43.\h3.45.\h3.51.\h3.57.\h3.59.\h3.63.\h3.65.
\h3.66.\h3.67.\h3.69.\h3.71.\h3.72.\h3.73.\h3.75.\h3.76.\h3.77.\h3.78.
\h3.79.\h3.81.\h3.83.\h3.84.\h3.85.\h3.87.\h3.89.\h3.91.\h3.93.\h3.95.
\h3.99.\h3.103.\h3.105.\h3.107.\h3.111.\h3.115.\h3.119.\h3.123.\h3.127.
\h3.131.\h3.141.\h3.165.                        
\h4.28.\h4.34.\h4.36.\h4.37.\h4.40.
\h4.42.\h4.44.\h4.46.\h4.48.\h4.49.\h4.50.\h4.52.\h4.54.\h4.55.\h4.56.
\h4.57.\h4.58.\h4.59.\h4.60.\h4.61.\h4.62.\h4.63.\h4.64.\h4.65.\h4.66.
\h4.67.\h4.68.\h4.69.\h4.70.\h4.71.\h4.72.\h4.73.\h4.74.\h4.76.\h4.77.
\h4.78.\h4.79.\h4.80.\h4.81.\h4.82.\h4.84.\h4.85.\h4.86.\h4.88.\h4.89.
\h4.90.\h4.91.\h4.92.\h4.93.\h4.94.\h4.96.\h4.97.\h4.98.\h4.100.\h4.101.
\h4.102.\h4.104.\h4.106.\h4.108.\h4.109.\h4.110.\h4.112.\h4.114.\h4.116.
\h4.118.\h4.120.\h4.121.\h4.122.\h4.124.\h4.126.\h4.128.\h4.130.\h4.136.
\h4.142.\h4.144.\h4.148.\h4.154.\h4.162.\h4.166.
\h5.20.\h5.29.\h5.31.\h5.33.\h5.35.\h5.37.\h5.38.\h5.39.
\h5.40.\h5.41.\h5.42.\h5.43.\h5.44.\h5.45.\h5.46.\h5.47.\h5.48.\h5.49.
\h5.50.\h5.51.\h5.52.\h5.53.\h5.54.\h5.55.\h5.56.\h5.57.\h5.58.\h5.59.
\h5.60.\h5.61.\h5.62.\h5.63.\h5.64.\h5.65.\h5.66.\h5.67.\h5.68.\h5.69.
\h5.70.\h5.71.\h5.72.\h5.73.\h5.74.\h5.75.\h5.76.\h5.77.\h5.78.\h5.79.
\h5.80.\h5.81.\h5.82.\h5.83.\h5.84.\h5.85.\h5.86.\h5.87.\h5.88.\h5.89.
\h5.90.\h5.91.\h5.92.\h5.93.\h5.94.\h5.95.\h5.97.\h5.98.\h5.99.\h5.101.
\h5.103.\h5.104.\h5.105.\h5.107.\h5.108.\h5.109.\h5.110.\h5.111.\h5.113.
\h5.115.\h5.116.\h5.117.\h5.119.\h5.121.\h5.122.\h5.123.\h5.125.\h5.127.
\h5.128.\h5.129.\h5.131.\h5.133.\h5.135.\h5.137.\h5.139.\h5.141.\h5.143.
\h5.145.\h5.149.\h5.153.\h5.161.\h5.165.
\h6.26.\h6.27.\h6.28.\h6.30.\h6.32.
\h6.33.\h6.34.\h6.36.\h6.37.\h6.38.\h6.39.\h6.40.\h6.41.\h6.42.\h6.43.
\h6.44.\h6.45.\h6.46.\h6.47.\h6.48.\h6.49.\h6.50.\h6.51.\h6.52.\h6.53.
\h6.54.\h6.55.\h6.56.\h6.57.\h6.58.\h6.59.\h6.60.\h6.61.\h6.62.\h6.63.
\h6.64.\h6.65.\h6.66.\h6.67.\h6.68.\h6.69.\h6.70.\h6.71.\h6.72.\h6.73.
\h6.74.\h6.75.\h6.76.\h6.77.\h6.78.\h6.79.\h6.80.\h6.81.\h6.82.\h6.83.
\h6.84.\h6.85.\h6.86.\h6.87.\h6.88.\h6.89.\h6.90.\h6.91.\h6.92.\h6.93.
\h6.94.\h6.95.\h6.96.\h6.97.\h6.98.\h6.99.\h6.100.\h6.101.\h6.102.\h6.103.
\h6.104.\h6.105.\h6.106.\h6.107.\h6.108.\h6.109.\h6.110.\h6.111.\h6.112.
\h6.114.\h6.115.\h6.116.\h6.117.\h6.118.\h6.120.\h6.122.\h6.124.\h6.126.
\h6.127.\h6.128.\h6.132.\h6.136.\h6.138.\h6.140.\h6.142.\h6.144.\h6.148.
\h6.150.\h6.152.\h6.153.\h6.156.\h6.160.\h6.162.\h6.164.\h6.168.
\h7.19.\h7.23.\h7.25.\h7.27.\h7.28.\h7.29.\h7.30.
\h7.31.\h7.32.\h7.33.\h7.34.\h7.35.\h7.36.\h7.37.\h7.38.\h7.39.\h7.40.
\h7.41.\h7.42.\h7.43.\h7.44.\h7.45.\h7.46.\h7.47.\h7.48.\h7.49.\h7.50.
\h7.51.\h7.52.\h7.53.\h7.54.\h7.55.\h7.56.\h7.57.\h7.58.\h7.59.\h7.60.
\h7.61.\h7.62.\h7.63.\h7.64.\h7.65.\h7.66.\h7.67.\h7.68.\h7.69.\h7.70.
\h7.71.\h7.72.\h7.73.\h7.74.\h7.75.\h7.76.\h7.77.\h7.78.\h7.79.\h7.80.
\h7.81.\h7.82.\h7.83.\h7.84.\h7.85.\h7.86.\h7.87.\h7.88.\h7.89.\h7.90.
\h7.91.\h7.92.\h7.93.\h7.94.\h7.95.\h7.96.\h7.97.\h7.98.\h7.99.\h7.100.
\h7.101.\h7.102.\h7.103.\h7.104.\h7.105.\h7.106.\h7.107.\h7.108.\h7.109.
\h7.110.\h7.111.\h7.112.\h7.113.\h7.114.\h7.115.\h7.116.\h7.117.\h7.119.
\h7.120.\h7.121.\h7.122.\h7.123.\h7.124.\h7.125.\h7.127.\h7.130.\h7.131.
\h7.133.\h7.135.\h7.137.\h7.139.\h7.140.\h7.141.\h7.142.\h7.143.\h7.145.
\h7.147.\h7.149.\h7.151.\h7.153.\h7.154.\h7.155.\h7.157.\h7.159.\h7.160.
\h7.161.\h7.163.\h7.167.\h7.169.
\h8.22.\h8.24.\h8.25.\h8.26.\h8.27.\h8.28.\h8.29.\h8.30.
\h8.31.\h8.32.\h8.33.\h8.34.\h8.35.\h8.36.\h8.37.\h8.38.\h8.39.\h8.40.
\h8.41.\h8.42.\h8.43.\h8.44.\h8.45.\h8.46.\h8.47.\h8.48.\h8.49.\h8.50.
\h8.51.\h8.52.\h8.53.\h8.54.\h8.55.\h8.56.\h8.57.\h8.58.\h8.59.\h8.60.
\h8.61.\h8.62.\h8.63.\h8.64.\h8.65.\h8.66.\h8.67.\h8.68.\h8.69.\h8.70.
\h8.71.\h8.72.\h8.73.\h8.74.\h8.75.\h8.76.\h8.77.\h8.78.\h8.79.\h8.80.
\h8.81.\h8.82.\h8.83.\h8.84.\h8.85.\h8.86.\h8.87.\h8.88.\h8.89.\h8.90.
\h8.91.\h8.92.\h8.93.\h8.94.\h8.95.\h8.96.\h8.97.\h8.98.\h8.99.\h8.100.
\h8.101.\h8.102.\h8.103.\h8.104.\h8.105.\h8.106.\h8.107.\h8.108.\h8.109.
\h8.110.\h8.111.\h8.112.\h8.113.\h8.114.\h8.115.\h8.116.\h8.118.\h8.119.
\h8.120.\h8.121.\h8.122.\h8.123.\h8.124.\h8.125.\h8.126.\h8.128.\h8.130.
\h8.131.\h8.132.\h8.134.\h8.136.\h8.137.\h8.138.\h8.140.\h8.142.\h8.143.
\h8.144.\h8.146.\h8.148.\h8.149.\h8.150.\h8.151.\h8.152.\h8.154.\h8.155.
\h8.156.\h8.158.\h8.160.\h8.161.\h8.162.\h8.163.\h8.164.\h8.166.\h8.167.
\h8.168.\h8.170.
\h9.19.\h9.21.\h9.22.\h9.23.\h9.24.\h9.25.\h9.26.\h9.27.\h9.28.\h9.29.
\h9.30.\h9.31.\h9.32.\h9.33.\h9.34.\h9.35.\h9.36.\h9.37.\h9.38.\h9.39.
\h9.40.\h9.41.\h9.42.\h9.43.\h9.44.\h9.45.\h9.46.\h9.47.\h9.48.\h9.49.
\h9.50.\h9.51.\h9.52.\h9.53.\h9.54.\h9.55.\h9.56.\h9.57.\h9.58.\h9.59.
\h9.60.\h9.61.\h9.62.\h9.63.\h9.64.\h9.65.\h9.66.\h9.67.\h9.68.\h9.69.
\h9.70.\h9.71.\h9.72.\h9.73.\h9.74.\h9.75.\h9.76.\h9.77.\h9.78.\h9.79.
\h9.80.\h9.81.\h9.82.\h9.83.\h9.84.\h9.85.\h9.86.\h9.87.\h9.88.\h9.89.
\h9.90.\h9.91.\h9.92.\h9.93.\h9.94.\h9.95.\h9.96.\h9.97.\h9.98.\h9.99.
\h9.100.\h9.101.\h9.102.\h9.103.\h9.104.\h9.105.\h9.106.\h9.107.\h9.108.
\h9.109.\h9.110.\h9.111.\h9.112.\h9.113.\h9.114.\h9.115.\h9.116.\h9.117.
\h9.118.\h9.119.\h9.120.\h9.121.\h9.122.\h9.123.\h9.124.\h9.125.\h9.126.
\h9.127.\h9.128.\h9.129.\h9.130.\h9.131.\h9.132.\h9.133.\h9.134.\h9.135.
\h9.136.\h9.137.\h9.138.\h9.139.\h9.141.\h9.143.\h9.144.\h9.145.\h9.146.
\h9.147.\h9.148.\h9.149.\h9.150.\h9.151.\h9.153.\h9.154.\h9.155.\h9.156.
\h9.157.\h9.158.\h9.159.\h9.161.\h9.162.\h9.163.\h9.165.\h9.167.\h9.168.
\h9.169.                        \h10.18.\h10.20.\h10.22.\h10.23.
\h10.24.\h10.25.\h10.26.\h10.27.\h10.28.\h10.29.\h10.30.\h10.31.
\h10.32.\h10.33.\h10.34.\h10.35.\h10.36.\h10.37.\h10.38.\h10.39.
\h10.40.\h10.41.\h10.42.\h10.43.\h10.44.\h10.45.\h10.46.\h10.47.
\h10.48.\h10.49.\h10.50.\h10.51.\h10.52.\h10.53.\h10.54.\h10.55.
\h10.56.\h10.57.\h10.58.\h10.59.\h10.60.\h10.61.\h10.62.\h10.63.
\h10.64.\h10.65.\h10.66.\h10.67.\h10.68.\h10.69.\h10.70.\h10.71.
\h10.72.\h10.73.\h10.74.\h10.75.\h10.76.\h10.77.\h10.78.\h10.79.
\h10.80.\h10.81.\h10.82.\h10.83.\h10.84.\h10.85.\h10.86.\h10.87.
\h10.88.\h10.89.\h10.90.\h10.91.\h10.92.\h10.93.\h10.94.\h10.95.
\h10.96.\h10.97.\h10.98.\h10.99.\h10.100.\h10.101.\h10.102.\h10.103.
\h10.104.\h10.105.\h10.106.\h10.107.\h10.108.\h10.109.\h10.110.\h10.111.
\h10.112.\h10.113.\h10.114.\h10.115.\h10.116.\h10.117.\h10.118.\h10.119.
\h10.120.\h10.121.\h10.122.\h10.123.\h10.124.\h10.125.\h10.126.\h10.127.
\h10.128.\h10.129.\h10.130.\h10.131.\h10.132.\h10.133.\h10.134.\h10.135.
\h10.136.\h10.137.\h10.138.\h10.139.\h10.140.\h10.141.\h10.142.\h10.143.
\h10.144.\h10.145.\h10.146.\h10.148.\h10.149.\h10.150.\h10.151.\h10.152.
\h10.153.\h10.154.\h10.155.\h10.156.\h10.157.\h10.158.\h10.159.\h10.160.
\h10.162.\h10.163.\h10.164.\h10.166.\h10.167.\h10.168.\h10.169.\h10.170.
\end{picture}}}

\newcommand{\ttbs}{\char'134}
\newcommand{\AmS}{{\protect\the\textfont2
  A\kern-.1667em\lower.5ex\hbox{M}\kern-.125emS}}

\title{Strings on Calabi--Yau spaces and Toric Geometry}

\author{Maximilian Kreuzer\address{Institut f\"ur Theoretische Physik, 
        Technische Universit\"at Wien\\
        Wiedner Hauptstra\ss e 8--10, A-1040 Wien, AUSTRIA}%
        \thanks{This work is supported in part by the Austrian Research Funds 
        FWF under grant Nr. P14639-TPH.}}

\begin{document}

\begin{abstract}
After a brief introduction into the use of Calabi--Yau varieties in
string dualities, and the role of toric geometry in that context,
we review the classification of toric Calabi-Yau hypersurfaces
and present some results on complete intersections. While no proof of
the existence of a finite bound on the Hodge numbers is known, all
new data stay inside the familiar range $h_{11}+h_{12}\le 502$. 
\vspace{1pc}
\end{abstract}

\maketitle

\section{INTRODUCTION}

Calabi--Yau manifolds are an important ingredient for constructing 
(quasi-)realistic string models, because $N=1$ supersymmetry below the string
scale essentially implies a complex structure and the existence of a Ricci-flat
Kähler metric on the effective space-time manifold \cite{ca85,gr98}. 
While exact conformal techniques provide an important complementary tool 
\cite{sc90,fo90}, geometric compactification has the advantage that we can 
vary the moduli of a string model and are not stuck on isolated 
``Gepner points'', where the conformal field theory is rational.

The most important problem of string theory, its lack of
a non-perturbative definition, is still unsolved. Nevertheless, 
some aspects of non-perturbative string physics became accessible in 1995
through the discovery of nonperturbative dualities \cite{wi95,ka95,po95}. 
These allow us to compute certain quantities at strong coupling by relating 
them to a weak coupling situation in a dual model. These dualities have been
tested, for example, by comparing quantities that are protected against 
nonperturbative corrections by a sufficient amount of supersymmetry.

Dualities exchange elementary degrees of freedom with composite
(solitonic) states that become light at strong coupling. In perturbative
string theory the light particle states come from string oscillations, 
Kaluza-Klein excitations, and winding modes. The latter can be regarded as
$\sigma$ model solitons and lead, for example, to 
T-duality of torus compactifications \cite{Tdual}.
A generalization of this duality to curved space, called 
mirror symmetry \cite{lvw,st96}, operates between topologically distinct 
Calabi--Yau manifolds by exchanging complex structure and Kähler moduli
and allows us to sum up sigma model corrections to certain quantities 
\cite{ca91,as94,mo95}.

Contributions to the light particle spectrum that are non-perturbative in
the string coupling essentially come from two different sources: Solitonic
branes
that wrap small cycles of a compact internal manifold become massless in 
the limit where the volume of the cycle vanishes. The additional massless
particles that arise in this way turned out to match the change in the
Hodge data that occurs in conifold transitions \cite{ca89} between Calabi--Yau
manifolds of different topologies via singular limits. 
In type II compactifications the physics of such a transition
is smooth, with D-branes wrapped on one side of the singularity 
turning into elementary particles \cite{st95,gr95}. 
Since the D-branes can be identified with
black supergravity branes \cite{po95}, 
this effect was called black hole condensation.

The second type of contribution arises when the background itself is
non-perturbative. Then, for example, open strings stretching between 
nearby D-branes lead to gauge and matter fields that contributes to the low 
energy effective theory \cite{po96,ka96,be96}, but are 
localized in the internal 
dimensions (which could become quite large as compared to the Planck scale).
Witten observed that strongly coupled IIA strings grow an additional effective
dimension. The resulting limit of string theory, whose effective low energy 
theory is eleven-dimensional supergravity, is called M-theory \cite{wi95}.
IIB strings, on the other hand, are selfdual with duality group $SL(2,\IZ)$,
whose geometrization led Vafa to propose a twelve-dimensional origin of
this duality, called
F-theory \cite{va96}. In many of the resulting dualities fibration structures 
play an important role \cite{vamo,Arev} and toric geometry provides us with 
a very general and powerful framework for analyzing the relevant mathematical
structures.

We begin with some general comments on constructing Calabi--Yau manifolds.
In section 3 we introduce toric geometry and comment on topology change. 
Then we come to the classification program of reflexive polyhedra, which 
naturally encode the combinatorial data that define toric Calabi--Yau 
hypersurfaces. We close with our results and discuss their use
for analyzing string dualities. 

\section{HOW TO MAKE A CALABI--YAU}

Compact Kähler spaces with $c_1=0$ are usually constructed as (intersections 
of) hypersurfaces of some simpler compact spaces, like projective space 
$\IP^n$, Grassmannians, or some more general coset spaces. These symmetric 
ambient spaces all have positive Ricci curvature, whose contribution to 
$c_1$ must be compensated by the hypersurface equations. The prototype of 
this construction is the quintic in $\IP^4$, which is defined as the 
vanishing locus of a homogeneous polynomial of degree 5. A generalization 
to complete intersections in products of projective spaces produced 265 
different pairs of Betti numbers $b_2=h_{11}$ and $b_3=2h_{12}+2$ 
(cf. {\em Fig.~1})
\cite{gr87,CICY}. 
Here $h_{11}$ counts the Kähler moduli, which mostly correspond to the
volumes of the ambient space factors. The complex structure 
deformations $h_{12}$, on the other hand,
get a much larger contribution from the parameters in the defining equations. 
Thus all Euler numbers $\chi=2(h_{11}-h_{12})$ turned out negative.

\begin{figure}
\centerline{\BP(120,65)(-55,-20)\unitlength=3pt
          \put(0,12)1\put(5,8)0\put(10,4)0\put(0,-12)1\put(15,0)1\put(-15,0)1
        \put(-5,-8)0\put(-10,-4)0\put(5,-8)0\put(-10,4)0\put(-5,8)0
        \put(10,-4)0\put(-5,0){$\!\!h_{12}$}\put(5,0){$\!\!h_{12}$}
        \put(0,-4){$\!\!h_{11}$}\put(0,4){$\!\!h_{11}$}\EP}
\caption{The Hodge diamond of a Calabi--Yau manifold. The Betti numbers
        are $b_n=\sum\limits_{p+q=n} h_{pq}$.} 
\end{figure}

This seemed to be bad news for the idea of mirror symmetry, which exchanges
$h_{11}$ and $h_{12}$, but
from the 
conformal field theory point of view simply corresponds to charge conjugation
\cite{lvw}.
Using, however, weighted projective spaces as ambient spaces, the resolution 
of the singularities that come from the weighted identification
\begin{equation}
        (z_1,\ldots,z_N)\sim(\lambda^{q_1}z_1,\ldots,\lambda^{q_N}z_N)
\end{equation}
introduces additional Kähler moduli, and indeed, a construction of some 6000
weight systems $\vec q=(q_1,\ldots,q_N)$ that admit Calabi--Yau hypersurfaces
produced a set of Hodge data that was 90\% mirror symmetric \cite{CLS}.
Surprisingly, a complete classification of all 7555 transversal
weights made the result less symmetric, with only 83\% of the data coming in
mirror pairs \cite{nms,KlS}, and generalizations like orbifolding \cite{aas}
and discrete torsion \cite{dt} did not improve the situation. In retrospective,
this is a consequence of the fact that the orbifolding construction of mirror
manifolds discovered by Berglund and Hübsch \cite{be93,mmi,odt,ade} only 
works for certain subclasses of transversal weights, and many more 
complicated weights only showed up in the complete lists.

Progress in the understanding of mirror symmetry came via a further 
generalization of the ambient spaces under consideration to toric varieties. 
This development is due to Batyrev, who found a manifestly mirror symmetric
construction of toric Calabi--Yau spaces for which the mirror map is 
implemented as a simple combinatorial duality of lattice polyhedra 
\cite{bat}.

\section{TORIC GEOMETRY}

The simplest approach to toric varieties \cite{Ful,Oda} uses the homogeneous
coordinate ring that has been introduced by Cox \cite{Cox1}.
This construction is similar to weighted 
projective space, but with an arbitrary number $n$ of scaling identifications 
\begin{equation}
(z_1,\ldots,z_N)\sim (\lambda^{q_1^{(i)}}z_1,\ldots,\lambda^{q_N^{(i)}}z_N), 
        ~~~     i\le n.
\end{equation}
The rational scaling weights $\vec q\,{}^{(i)}$ can be encoded by linear 
relations of lattice vectors as shown in {\em Fig. 2}. 
In order for the resulting quotient $(\IC^N-Z)/(\IC^*)^n$ to be well behaved,
it is necessary to subtract a more general exceptional set $Z$, which is 
defined in terms of a collection $\Sigma$ of cones $\sigma$ that is closed 
under intersections (the ``algebraic torus'' $\IC^*$ is the multiplicative 
group of non-zero complex numbers). 
$\Sigma$ is called the fan of the toric variety and
the exceptional set $Z$ ensures that only coordinates whose respective
vectors belong to the same cone can vanish simultaneously.

\begin{figure}
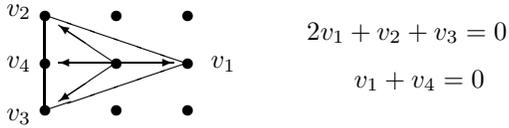
\unitlength=1.8pt         
\BP(50,25)(-27,-5)     
\def\putvec#1,#2,#3,#4,#5){\put#1,#2){\vector(#3,#4){#5}}}
\put(50,-5){$v_1+v_4=0$}\put(40,5){$2v_1+v_2+v_3=0$} \def\lila{\TC{0 1 0 0}}
\lila   \put(-.1,0){\drawline(15,0)(-15,10)\drawline(15,0)(-15,-10)
        \drawline(-15,10)(-15,-10)}
        %
\blue   \put(-.1,0){
                        \putvec(0,0,1,0,12)\put(-23,-12){$v_3$}
        \putvec(0,0,-3,-2,12)\putvec(0,0,-3,2,12)\putvec(0,0,-1,0,12)
        \put(20,-1){$v_1$}\put(-23,-1){$v_4$}\put(-23,10){$v_2$}
        }
\black  {
\matrixput(-15,-10)(15,0)3(0,10)3{\circle*2}}
\EP
\caption{The  scaling weights for the coordinates $z_i$ are
        $\vec q^{~(1)}=(2,1,1,0)$ and $\vec q^{~(2)}=(1,0,0,1)$. 
        $z_4$ implements the blowup of the singular point (1:0:0), 
        which gets excluded from $W\IP_{2,1,1}^2$ by adding $v_4$.}
\end{figure}

In the example of {\em Fig. 2} the vectors $\{v_1,v_2,v_3\}$ define the
weighted projective space $W\IP_{2,1,1}^2$ with scaling weights 
$\vec q=(2,1,1)$. The exceptional set would only consist of the origin, as any
subset of two coordinates belongs to one of the three cones spanned by
$(v_i,v_j)$. It is easy to see that $W\IP_{2,1,1}^2$ has a $\IZ_2$ quotient 
singularity at the point with homogeneous coordinates $(1:0:0)$. This 
singularity can be resolved be introducing the additional vector $v_4$ and
the corresponding additional scaling relation. 
$v_4$ subdivides the cone $(v_2,v_3)$ and thereby moves the singular point 
to the new exceptional set $Z=\{z_2=z_3=0\}\cup\{z_1=z_4=0\}$.
The singularity gets blown up to a complete $\IP^1$: Whenever $z_4\neq0$ we 
can scale it to 1 and recover all of $W\IP_{2,1,1}^2$ except for 
the point $z_2=z_3=0$.
For $z_4=0$ we can scale away $z_1$, but are left with a complete $\IP^1$ that 
is parametrized by the homogeneous coordinates $(z_2:z_3)$.             
It can be shown that a toric 
variety is non-singular if all cones are simplicial and regular (i.e. they are
generated by vectors that span a simplex of volume 1 in lattice units)
\cite{Ful}. This is the case for the fan shown in {\em Fig. 2}, so that the
blowup of the singular point desingularizes $W\IP_{2,1,1}^2$.

\begin{figure}
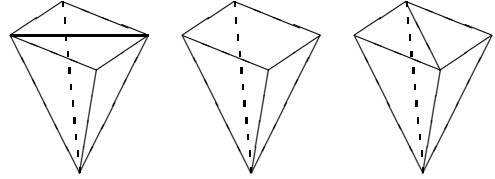
\unitlength=1.3pt
\def\ConeSigma{\drawline(0,0)(-20,40)\drawline(0,0)(20,40)
        \dashline2(0,0)(-5,50)\drawline(0,0)(5,30)
        \drawline(5,30)(20,40)\drawline(5,30)(-20,40)
        \drawline(-5,50)(20,40)\drawline(-5,50)(-20,40)}
\BP(100,52)(0,20)
\put(30,10){\ConeSigma
        \put(0,0){\blue\linethickness{2pt}\drawline(-20,40)(20,40)}\black}
\put(80,10){\ConeSigma  }
\put(130,10){\ConeSigma
        \put(0,0){\blue\linethickness{2pt}\drawline(-5,50)(5,30)}\black}
\EP
\caption{Flop transition: Different triangulations of a cone in the fan
        lead to topologically destinct desingularizations.}
\end{figure}

The dependency of the exceptional set $Z$ on the complete fan (and not just on
the generators of the 1-dimensional cones, which define the homogeneous
coordinates) commences only in 3 dimensions: In {\em Fig. 3} we observe
that a quadratic cone can be triangulated in two different ways, thereby
increasing the exceptional set in two different ways and yielding two 
different blowups of the singularity of the non-simplicial situation.
The respective toric varieties are topologically distinct, as they have
different intersection numbers, but they
turn out to have identical Hodge numbers. Connecting topologically
distinct spaces through this relatively mild kind of singularity is called
a flop transition, and this was the first setting where it could be shown
that the corresponding physics varies smoothly despite of the topology change
\cite{as94}.

\section{REFLEXIVE POLYHEDRA}

\begin{figure*}[htb]
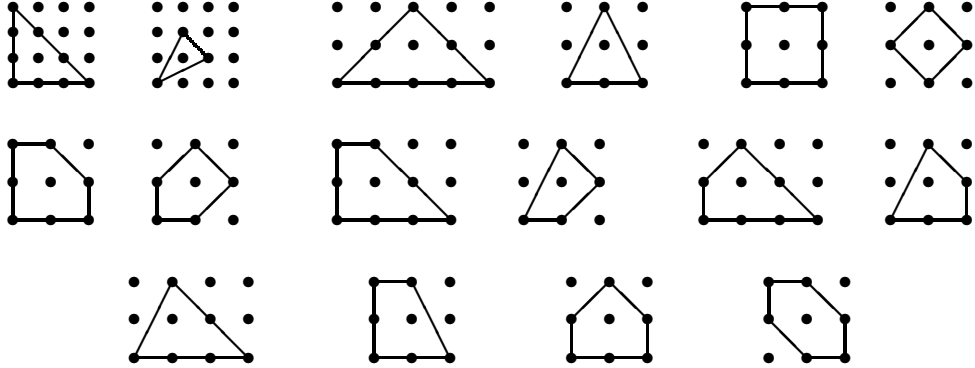

\vspace*{-9pt}\TwoDimRP \vspace*{-25pt}                 \label{TwoDimRP}
\caption{All 16 reflexive polygons in 2D: The first 3 dual pairs are 
        maximal/minimal and contain all others as subpolygons, while the last 
        4 polygons are selfdual.}
\end{figure*}

The lattice supporting the cone $\Sigma$ of a toric variety $\IP_\Sigma$
is usually called the $N$ lattice. The points $m$ of the dual lattice 
$M=\mathrm{Hom}(N,\IZ)$ also play an important role, because they correspond 
to monomials $\prod z_i^{\langle m,v_i\rangle}$ that provide sections of 
certain line bundles over $\IP_\Sigma$. These line bundles can be constructed 
in terms of lattice polyhedra $\Delta \subset M$. 
We define the dual of a polyhedron with $0\in\Delta$ as
\begin{equation}
        \Delta^*=\{y\in N_\IR~| ~ \langle y,x\rangle\ge-1~\forall 
        x\in\Delta\subset M_\IR\},
\end{equation}
where $M_\IR$ denotes the real extension $M\otimes_\IZ\IR$ of the lattice $M$.
The important result of Batyrev was that the generic section of a line 
bundle that corresponds to $\Delta$ 
defines a Calabi--Yau hypersurface in $\IP_\Sigma$ if $\Delta^*$ is a lattice 
polytope and if the fan $\Sigma$ is the fan of cones over the faces of 
$\Delta^*$ \cite{bat}
(to get a smooth Calabi--Yau manifold, this fan still has to be triangulated).
Moreover, the Hodge numbers can be computed by a simple combinatorial
formula in terms of the numbers of interior points of dual faces of 
the reflexive pair of polyhedra, and mirror symmetry is simply implemented by 
exchanging $\Delta$ and  $\Delta^*$.

A lattice polyhedron whose dual vertices belong to the dual lattice is
called reflexive. This condition is equivalent to $\Delta$ having exactly 
one interior point with all facets at distance 1 (i.e. there are no parallel
lattice hyperplanes between a facet and the interior point). The classification
of toric Calabi--Yau hypersurfaces therefore reduces to the 
classification of reflexive polyhedra.

\begin{figure*}[htb]
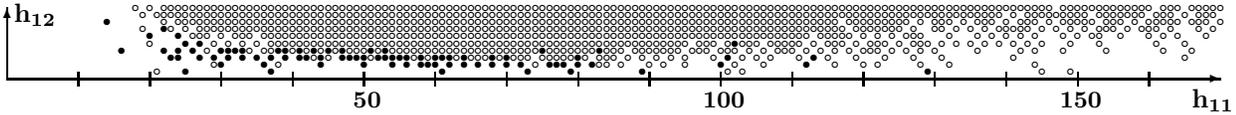
\CICYspec   \vspace{-22pt}          \label{CICYspec}
\caption{87 new CICY spectra in the background of hypersurfaces 
         with $h_{11}\le170$ and $h_{12}\le10$.}
\end{figure*}

The 16 reflexive polygons that exist
in two dimensions were first constructed by Batyrev and are shown in 
{\em Fig. 4} in order to illustrate our approach to the classification, which 
has now been completed in up to 4 dimensions \cite{crp,hs,k3,c3d,ams,c4d}.
The first step in this program is the identification of a set of objects that
contain all others as subpolytopes. In the two-dimensional case these are
the three polygons with 10, 8 and 9 points in the first line of {\em Fig. 4}.
Their duals are minimal in the sense that they loose the interior lattice
point if one of their vertices is dropped. The linear relations among the 
vertices of (a simplex decomposition of) these minimal objects are used as 
our starting point \cite{crp,ams} and we define the corresponding weight 
systems via the baricentric coordinates of the origin, i.e. $\sum q_iv_i=0$ 
($q_i$ can be chosen to be integer without common divisors). 

With $d=\sum q_i$
this gives an efficient (coordinate independent) description of the dual 
maximal object by the equation $\sum  q_in_i=d$ for non-negative integers 
$n_i\ge0$. 
If the unique candidate $n_i=1$ for an interior point is indeed in the interior
of the convex hull we call $\vec  q$ an ``IP weight system''. The 
maximal simplexes in {\em Fig. 4}, for example, correspond to $n_1+n_2+n_3=3$
and $n_1+n_2+2n_3=4$, respectively 
(the points $\vec n\in\Delta$ correspond to the Newton polyhedron
of a quasi-homogeneous polynomial, providing a link to
Calabi--Yau hypersurfaces in $W\IP^4$). 
The square, on the other hand, is the convex hull
of two one-dimensional simplexes and requires a combination of the two weight
systems (1,1,0,0) and (0,0,1,1). This results in the description of the dual
maximal square by $n_1+n_2=n_3+n_4=2$.

The constructive proof that the number of rational IP weight systems
is finite in principle provides an algorithm for the construction of all 
reflexive polytopes \cite{crp,hs,ams}. To make it work in finite time it 
still requires a number of refinements that were developed during our
implementations for the classification in 3 and 4 dimensions \cite{c3d,c4d}. 
An important subtle point in this context is the fact that the IP weight 
system fixes a unique coarsest lattice $N_{\rm coarse}$, which is generated 
by the 
vertices of the minimal polytope, but leaves open the possibility that the 
dual maximal object lives on a sublattice of the dual finest relevant lattice 
$M_{\rm fine}=\mathrm{Hom}(N_{\rm coarse},\IZ)$.

\section{RESULTS}

In three dimensions the 95 IP weight systems have been know for some time.
Adding the relevant combined IP weights we found 14 maximal polytopes
on $M_{\rm fine}$ and one additional maximal polytope on a $\IZ_2$ quotient 
of the lattice for the quartic in $\IP^3$. The total number of reflexive 
polytopes turned out to be 4319 \cite{c3d}.

In four dimensions there are 184026 IP weight systems and we had to begin
with 308 maximal reflexive objects on $M_{\rm fine}$. We found 25 additional
maximal reflexive polyhedra on sublattices and a total of 473800776 reflexive
subpolytopes, which completes the classification in 4 dimensions \cite{c4d}.
The corresponding toric Calabi--Yau hypersurfaces have 30108 different
pairs of Hodge numbers. Storing the results requires about 5 GB of disk space.
While these cannot be downloaded from the internet in a reasonable time it 
is possible to search our data base for polytopes with certain properties 
at our web page \cite{KScy}.

A by product of our classification is the proof that all reflexive polytopes
in 3 and 4 dimensions are connected by a chain of singular transitions. These
correspond to blowing up (or down) a number of divisors that correspond
to points of $\Delta^*$, which are added (or dropped) to connect the
polyhedra. The intermediate singularities are more severe than in 
the case of flop transitions because also the Euler numbers change.
This has been discussed in detail by Strominger et al. in 
the simplest case of a conifold transition \cite{st95,gr95}.

\begin{figure}
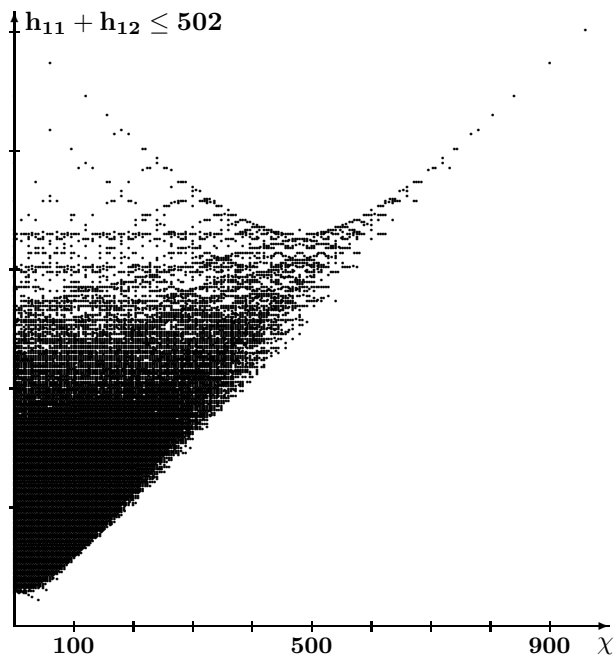
          \newcount\hsum \newcount\hdif
\unitlength=0.45pt

\caption{The 15122 (of 30108) hypersurface spectra with $h_{11}\le h_{12}$.
        The maximal value of $h_{11}+h_{12}$ comes from (251,251) and (491,11).}        
\end{figure}

Another important aspect of Calabi--Yau manifolds that is relevant for string
dualities are fibration structures. These are encoded in toric geometry in
a very transparent way because, for example, elliptic and K3 fibrations 
show up as reflexive section of $\Delta^*\subset N$ with dimensions two and
three, respectively. A large number of K3 fibrations has been found for the 
manifolds defined by the 184026 IP weight systems in 4 dimensions \cite{k3}.
Elliptic fibrations, which are important in F-theory, can be constructed
easily with certain types of combined weight systems, as has been discussed 
in \cite{ams,fft}.

While we have focused on toric hypersurfaces, the mirror involution was 
extended to complete intersection Calabi--Yau manifolds (CICY) by
Batyrev and Borisov \cite{bor,bb1,bb2,strh}, who again found a very 
beautiful combinatorial duality in terms of nef partitions, which are
certain decompositions of reflexive polyhedra into Minkowki sums. 
Using these results we analyzed a sizeable number of nef partitions of
5-dimensional polyhedra and computed the resulting Hodge numbers of
varieties of codimension 2 \cite{kr01}. The new spectra that we found are
shown in {\em Fig. 5}. In particular, we doubled the number of known spectra 
with $h_{11}=1$.

I would like to thank Dimitri Sorokin and his colleagues for perfectly
organizing this interesting conference, and Harald Skarke and Erwin 
Riegler for enjoyable collaborations.

\end{document}